\documentclass[sigconf]{acmart}
\AtBeginDocument{%
  \providecommand\BibTeX{{%
    \normalfont B\kern-0.5em{\scshape i\kern-0.25em b}\kern-0.8em\TeX}}}

\copyrightyear{2020}
\acmYear{2020}
\acmDOI{10.1145/3366423.3380083}

\acmConference[WWW '20]{Proceedings of The Web Conference 2020}{April 20--24, 2020}{Taipei, Taiwan}
\acmBooktitle{Proceedings of The Web Conference 2020 (WWW '20), April 20--24, 2020, Taipei, Taiwan} 
\acmPrice{}
\acmISBN{978-1-4503-7023-3/20/04}
\usepackage{verbatim}
\usepackage{amssymb}
\usepackage{bm}
\usepackage{amsmath}
\usepackage{subfigure}
\usepackage{longtable}
\usepackage{rotating}
\usepackage{multirow}
\usepackage{booktabs}
\usepackage{diagbox}

\begin{document}

\title{Structure-Feature based Graph Self-adaptive Pooling}

\author{Liang Zhang}
\authornotemark[1]
\email{liangzhang@xidian.edu.cn}
\affiliation{%
  \institution{Xidian University}
  \country{China}
}

\author{Xudong Wang}
\email{wxdssr233@163.com}
\affiliation{%
  \institution{Xidian University}
  \country{China}
}

\author{Hongsheng Li}
\email{hsli@stu.xidian.edu.cn}
\affiliation{%
  \institution{Xidian University}
  \country{China}
}

\author{Guangming Zhu}
\email{gmzhu@xidian.edu.cn}
\affiliation{%
  \institution{Xidian University}
  \country{China}
}

\author{Peiyi Shen}
\email{pyshen@xidian.edu.cn}
\affiliation{%
  \institution{Xidian University}
  \country{China}
}

\author{Ping Li}
\email{pli@bnc.org.cn}
\affiliation{%
  \institution{Shanghai BNC}
  \country{China}
}

\author{Xiaoyuan Lu}
\email{xylu@bnc.org.cn}
\affiliation{%
  \institution{Shanghai BNC}
  \country{China}
}

\author{Syed Afaq Ali Shah}
\email{Afaq.Shah@murdoch.edu.au}
\affiliation{%
  \institution{Murdoch University}
  \country{Australia}
}

\author{Mohammed Bennamoun}
\email{mohammed.bennamoun@uwa.edu.au}
\affiliation{%
  \institution{The University of Western Australia}
  \country{Australia}
}
\renewcommand{\shortauthors}{Liang Zhang and Xudong Wang, et al.}

\begin{abstract}
Various methods to deal with graph data have been proposed in recent years. However, most of these methods focus on graph feature aggregation rather than graph pooling. Besides, the existing top-k selection graph pooling methods have a few problems. \textbf{First}, to construct the pooled graph topology, current top-k selection methods evaluate the importance of the node from a single perspective only, which is simplistic and unobjective. \textbf{Second}, the feature information of unselected nodes is directly lost during the pooling process, which inevitably leads to a massive loss of graph feature information. To solve these problems mentioned above, we propose a novel graph self-adaptive pooling method with the following objectives: (1) to construct a reasonable pooled graph topology, structure and feature information of the graph are considered simultaneously, which provide additional veracity and objectivity in node selection; and (2) to make the pooled nodes contain sufficiently effective graph information, node feature information is aggregated before discarding the unimportant nodes; thus, the selected nodes contain information from neighbor nodes, which can enhance the use of features of the unselected nodes. Experimental results on four different datasets demonstrate that our method is effective in graph classification and outperforms state-of-the-art graph pooling methods.
\end{abstract}




\maketitle

\section{Introduction}
In the computer vision field, convolutional neural network (CNN) \cite{HG12}\cite{LiFF01}\cite{HeK16}\cite{Kri12}\cite{AK12} is a powerful tool for refining the information from images or videos for object recognition and relationship detection. However, in the real world, numerous data are organized graphically, such as protein-protein interaction and social networks \cite{LazerD09}\cite{NMK16}\cite{DDK15}. Although CNN is adaptive to process grid-structured data, such as images, it cannot easily deal with the non-Euclidean space data, such as graphs.

To date, many scholars have proposed graph convolution networks (GCNs) in the spectral perspective, such as GCN with Chebyshev expansion (ChebConv) \cite{MD16} and semi-supervised classification with GCN (GCNConv) \cite{TN17}. Other non-spectral graph neural networks (GNNs) have also been proposed, such as GraphSAGE \cite{WLH17} and Graph Attention Network (GAT) \cite{PV18}, which have a similar function as GCN. However, all of these GCNs focus on the graph information aggregation rather than the pooling mechanism. 

Pooling (downsampling) \cite{AK12}\cite{CYL18} plays an important role in CNN because it can reduce the amount of data and acceletare the calculation, which facilitates the design of deep CNN and obtains improved performance. In recent years, a few methods studied graph pooling to utilize the pooling mechanism in GCN, which is used to reduce the number of nodes and edges in the graph. Some of the graph pooling methods use clusters of nodes to generate the pooled graph topology, such as DiffPool \cite{RY18} and EigenPooling \cite{YM19}, where several nodes are combined to generate new nodes through the assignment matrix. Others are top-k selection methods, such as gPool \cite{GH19} and SAGPool \cite{JL19}, in which node features and local structural information are used to compute the importance of the node, and then top-k nodes are selected as the pooling results. The pooled graph topology is decided by the selected top-k graph nodes.

However, the existing top-k selection graph pooling methods face two problems. \textbf{First}, in generating the pooled graph topology, these methods do not explicitly consider the graph structure and node feature representation together. \textbf{Second}, regarding the feature representations of the pooled nodes, the original features of the selected nodes are used as a new feature, and the features of the unselected nodes are discarded. Therefore, substantial graph information is lost in the pooling process, which may be important information on the graph.

In this paper, we propose a graph self-adaptive pooling (GSAPool) method to address these problems. Regarding the graph topology generation, our method evaluates in multiple ways the importance of nodes in accordance with the local structure and the feature information of the nodes. Regarding the feature generation after the pooling nodes are chosen, the feature aggregating process is used to ensure that the feature representation of the pooled nodes contain sufficiently effective information from the graph.

\section{Related Work}
A graph is a data structure that can be represented by a set $\{V, E\}$, where $V$ is the set of nodes and $E$ is the set of edges. Generally, the adjacency matrix $A\in {R}^{N \times N}$ indicates which nodes are connected and which edges are attributed. Therefore, this matrix is used to describe the structural information of the graph. In addition, the feature matrix is used to represent the node feature representation of the graph.
\subsection{Graph Convolution}
GCNs \cite{JB14}\cite{SY18}\cite{LY19} can be divided into two domains: spectral and non-spectral methods. Spectral methods utilize Laplacian matrix as spectral operator to refine the convolution in Fourier field. ChebConv \cite{MD16} uses the Laplacian matrix directly as a convolution operator. Without eigenvector decomposition, the number of parameters is reduced and the calculation can be accelerated. GCNConv \cite{TN17} extends convolution to the data of graph structure, which can obtain better graph representation and performs well in the semi-supervised task. By contrast, non-spectral approaches are meant to work directly on graphs in which the central node aggregates features from the neighbor nodes layer by layer. GraphSAGE \cite{WLH17} generates node embeddings by aggregating node feature information in neighborhood. GAT \cite{PV18} uses attention mechanism and calculates the attention score of adjacent nodes as weight value for feature information in the aggregation process.
\begin{figure*}[h]
\center{\includegraphics[width=15.5cm,height=8.5cm]{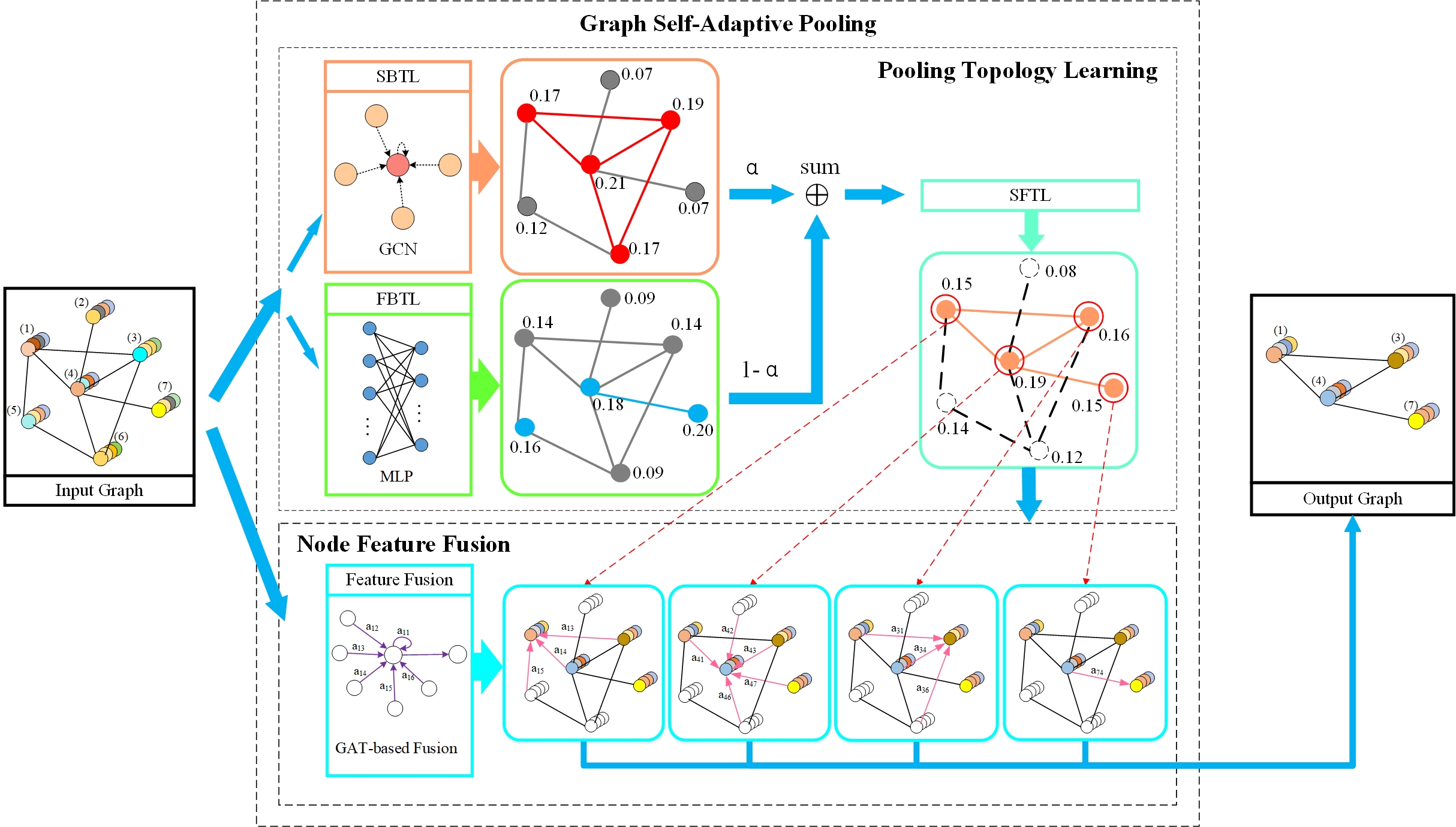}}
\caption{The illustration of the proposed GSAPool, which includes two parts: the pooling topology learning and the node feature fusion. The structure-based topology learning (SBTL) and the feature-based topology learning (FBTL) components in the first part calculate the importance scores according to the structure and feature information respectively. The structure-feature topology learning (SFTL) component fuses the scores to generate the final pooling topology. GAT-based node feature fusion is used in the second part to aggregate the features of the unselected nodes around the selected nodes.}
\label{figure1}
\end{figure*}
\subsection{Graph Pooling}
Pooling can enable CNNs to decrease the amount of parameters by scaling down the size of the input, which makes the training process highly efficient. Similar to CNN, the pooling mechanism can enable GNNs to reduce the parameters for better performance. From our perspective, the current pooling methods can be divided into two categories: cluster and top-k selection poolings. \\
\textbf{Cluster Pooling}: DiffPool \cite{RY18} and EigenPooling \cite{YM19} use cluster algorithms to decide the pooled graph topology, i.e., to select the new nodes in the pooling process. Assignment matrix is utilized to generate node clusters. In the assignment matrix, each column corresponds to the nodes of the graph before pooling, and the rows represent the nodes in the pooling results.
The concrete expression of the updating process for the adjacency and the feature matrices are denoted as follows:
$${X}^{(l+1)} = {S}^{{(l)}^T}{GNN(A, X)}^{(l)} \in R^{n_{l+1} \times d},\eqno{(1)}$$
$${A}^{(l+1)} = {S}^{{(l)}^T}A^{l}S^{(l)} \in R^{n_{l+1} \times n_{l+1}},\eqno{(2)}$$
where $S$ indicates the assignment matrix, $A$ is the adjacency matrix, and $X$ is the feature matrix. 

Similarly, in EigenPooling, the assignment matrix is used to update the graph information. The difference is that a pooling operator is utilized to refine the feature matrix. First, the graph is divided into several subgraphs in accordance with the assignment matrix. Then, the eigenvectors of the subgraphs are calculated by the Laplacian matrix as the pooling operator.\\

\textbf{Top-k Selection Pooling}:
SAGPool \cite{JL19} is a top-k selection pooling method in which the pooled graph topology is decided by the selected nodes. In SAGPool, GCNConv \cite{TN17} is used  to evaluate the importance of the nodes. Although GCNConv considers structure and feature information simultaneously, it uses the information implicitly. When obtaining the importance value of every node, several nodes with high scores are selected as the pooled graph.

In gPool \cite{GH19}, a learning vector maps the node feature into the importance scores. Then, the nodes are selected in accordance with the score. The following equation describes the calculation of the score:
$$ y = X^{l}p^{l}/||p^{l}||,\eqno{(3)}$$
where $y$ is the score vector that saves the feature scores of all nodes. $X^{l}$ is the feature matrix and $p^{l}$ is the learning vector of layer $l$.
Compared with SAGPool, gPool does not consider the graph structural information as a part of the importance score.

In cluster pooling algorithms, when generating the assignment matrix, the structure and feature information are used implicitly, thereby leading to the unreasonableness of the assignment matrix. In top-k selection pooling methods, the importance of nodes is considered from a single perspective, which is simplistic. Moreover, the features of the pooled nodes are still the original features, and the feature information of the unselected nodes is directly lost. \textbf{To address these problems, we consider that in generating the pooled graph topology, the node selection approaches should be diverse, and the feature information of the pooled nodes should include features of the adjacent nodes}. Therefore, we use additional evaluation standards for each node to generate highly accurate pooled graph topology. Furthermore, we use feature fusion method to enhance the feature representation ability of the pooled graph. As a result, our method of pooled graph generation is considerably diverse, objective, and accurate.

\section{Proposed Methods}
\subsection{Pooling Topology Learning}
The pooling topology learning method contains three parts: the structure-based topology learning (SBTL), the feature-based topology learning (FBTL), and the structure-feature topology learning (SFTL). The details of the learning method are illustrated in Figure \ref{figure1}.\\
\textbf{SBTL}: In general, a graph contains numerous nodes and edges, which indicate rich structural information. Therefore, it is effective to evaluate the importance of each node in accordance with its structural information. Given that GCNConv \cite{TN17} considers structural information, this method is suitable for evaluating the importance of a node. The expression of GCNConv is illustrated as follows:
$${S}_{1} = \sigma(\tilde{D}^{-\frac{1}{2}}\tilde A \tilde{D}^{-\frac{1}{2}}XW), \eqno{(4)}$$
where ${S}_{1}$ is the structural information score of a node calculated by GCNConv, $\tilde D$ is the sum of the degree and identity matrices of a graph, $\tilde A$ is composed of the adjency matrix and identity matrices, $X$ is the feature matrix and $W$ is the weight vector. $\sigma$ is the activate function often used in neural networks, such as \textit{tanh} and \textit{sigmoid}. 

As a complement, the score function kernel can be easliy replaced by other GNNs, such as ChebConv \cite{MD16}, GraphSAGE \cite{WLH17} and GAT \cite{PV18}, which can be expressed as follows:
$$S_1 = \sigma(GNN(A, X)), \eqno{(5)} $$
\textbf{FBTL}: In graph data, nodes usually contain feature information. Utilizing the node feature information for evaluation is important because a node can be largely represented by its feature. The effect of a node feature cannot be ignored.

We take a MLP \cite{JJH82} as the node feature extracting method because it is adaptive to aggregate feature information. The expression is as follows:
$$S_2 = \sigma(MLP(X)), \eqno{(6)}$$
where X is the node feature matrix and $\sigma$ is the activate function. $S_2$ is the aggregation of node feature, which can be regarded as node evaluation value. In accordance with the aggregation results, it can reserve additional important nodes.\\
\textbf{SFTL}: GCNConv \cite{TN17} is effective in refining the local structural information of a graph. Moreover, MLP \cite{JJH82} focuses on the feature information of a node. To make the standard of node evaluation highly objective and robust, we use the two methods to calculate the importance scores of the nodes. The diversity of the evaluation effectively increased by considering two different node evaluation methods synthetically, which reinforces the objectivity of the selection of final nodes. The combination is expressed as follows:
$${S}_{final} = \alpha{S}_{1}+(1-\alpha)S_2,\eqno{(7)}$$
The weight $\alpha$ is a user-defined hyperparameter. We sort the nodes in accordance with their scores, and use the top-k nodes as the pooling results.
\begin{figure}[h]
\center{\includegraphics[width=7.5cm, height=7cm]{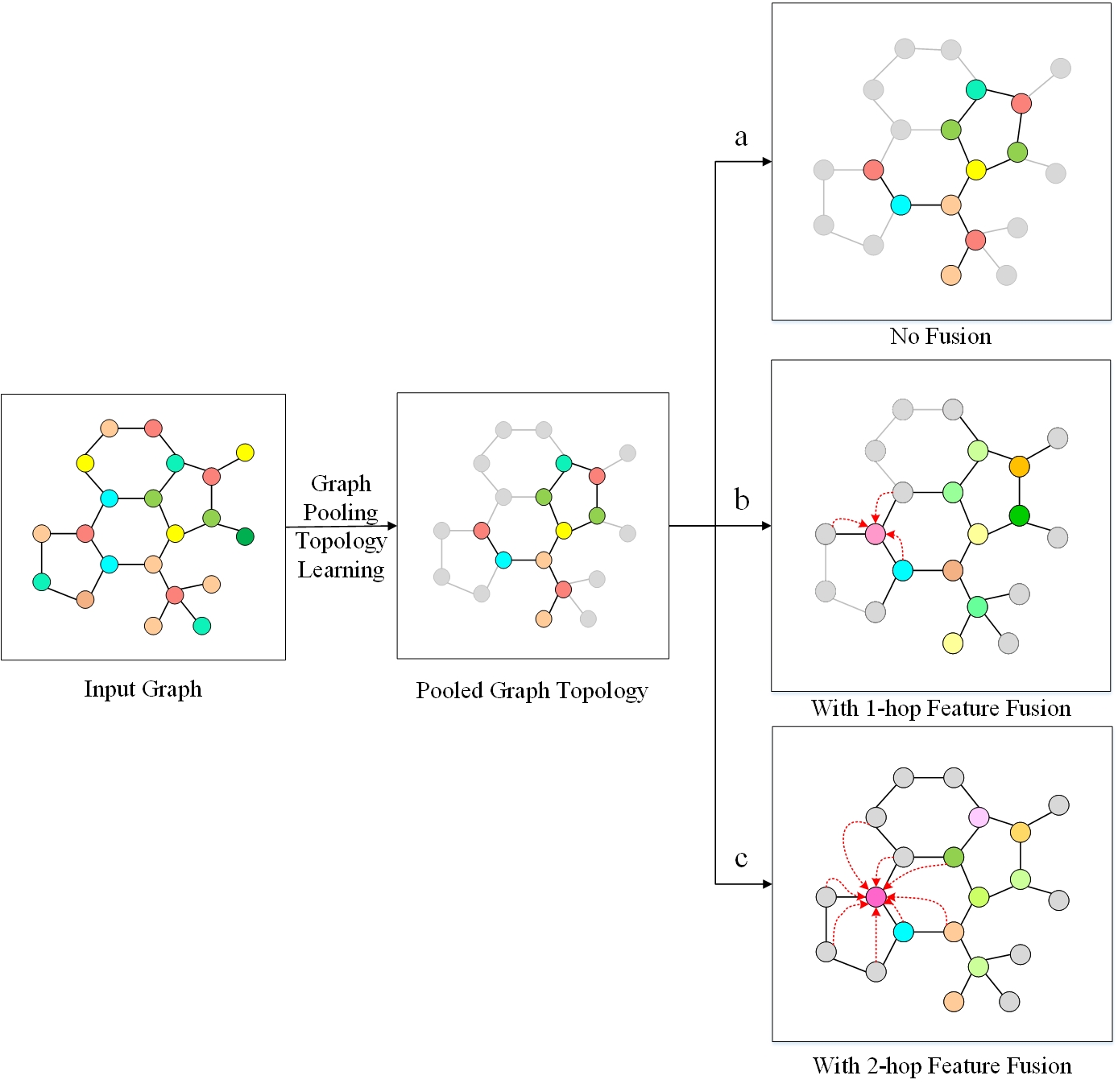}}
\caption{Different node feature fusion strategies: (a) the pooled nodes only reserve their own features, (b) the pooled nodes aggregate features within 1-hop neighbor nodes, (c) the pooled nodes aggregate features within n-hop neighbor nodes. The edge with arrow shows the direction of feature information flow in the fusion process.}
\label{figure2}
\end{figure}
\subsection{Pooled Node Feature Representation}
In the top-k selection pooling methods, only parts of nodes are selected as the pooling results. To use the information of the unselected nodes, we have to aggregate the features of the nodes before discarding them. The feature information of the nodes can be used more sufficiently, which makes the final graph embedding vector more representative. Figure \ref{figure2} shows the details of the feature fusion.

Two aggregation functions, i.e.,  the GCNConv \cite{TN17} and GAT \cite{PV18}, are evaluated in this study. The GAT is expressed as follows,
$${h_i}' = \sigma(\sum_{j \in N_i}a_{ij}^{k}W^k h_j),\eqno{(8)}$$
where $h_i$ is the feature vector of node $i$ and $h_j$ represents the neighbor nodes of node $i$. $N_i$ is the number of the adjacent nodes of node $i$. $a_{ij}^{k}$ is the attention value between $h_i$ and $h_j$ and $W$ is the weight matrix.

With the help of these fusion methods, the selected nodes can carry the feature information from the neighbor nodes. In this manner, the pooling result can highly represent the entire graph, as proven by our experimental results.
\begin{figure}[h]
\center{\includegraphics[width=4cm, height=6cm]{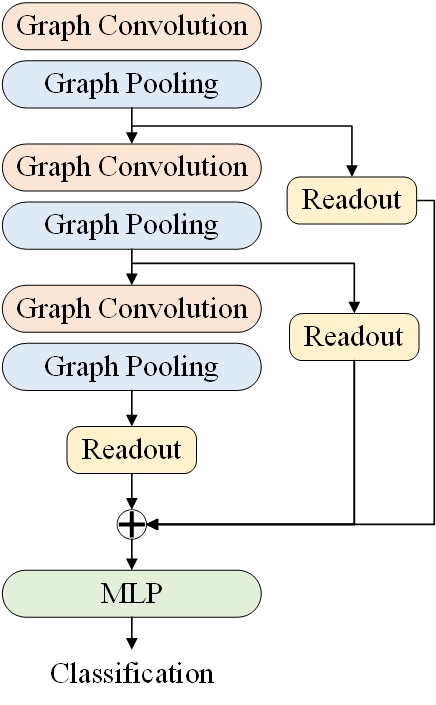}}
\caption{The illustration of the model architecture \cite{JL19} used in the proposed GSAPool.}
\label{figure3}
\end{figure}
\subsection{Model Architecture}
For a fair comparison, the model architecture of SAGPool \cite{JL19} is adopted in our experiments. All the comparative methods and our method use the same architectures. Figure \ref{figure3} shows the details of the model architecture.

\section{Experiments}

\subsection{Datasets} 
To verify whether our proposed method is effective, four graph datasets \cite{KK16} are used in the experiments. DD \cite{DPD13}\cite{SN11} contains graphs representing protein structures and the node is an amino acid. The graph labels indicate whether the protein is an enzyme. NCI-1 \cite{WN08}\cite{SN11} and NCI-109 \cite{WN08}\cite{SN11} are biological datasets used for anticancer activity classification. Every graph represents a chemical compound structure in which nodes and edges correspond to atoms and chemical bonds. Mutagenicity \cite{JK05}\cite{KR08} is a dataset that contains compounds used for medicine. The label of the compound indicates whether it has mutagenicity attributes. Table \ref{table1} summarizes the statistics of the datasets.
\begin{table*}[h]
\caption{Statistics of the experimental datasets.}
\begin{tabular}{lcccc}
\toprule
\textbf{Datasets} & \textbf{Number of Graphs} & \textbf{Number of  Classes} & \textbf{Average Number of Nodes} & \textbf{Average Number of Edges} \\
\toprule
DD & 1178 & 2 & 284.32 & 715.66\\
NCI-1 & 4110 & 2 & 29.87 & 32.30\\
NCI-109 & 4127 & 2 & 29.68 & 32.13\\
Mutagenicity & 4337 & 2 & 30.32 & 30.77\\
\bottomrule
\end{tabular}
\label{table1}
\end{table*}

\begin{table*}[h]
\caption{The evaluation of different pooling ratios based on the ${GSAPool}_{GCNConv+{Fusion}_{GAT}}$ architecture.}
\begin{tabular}{c|cccc}
\toprule
\multirow{2}{*}{\textbf{Pooling Ratio}} & \multicolumn{4}{c}{\textbf{Datasets}}\\
\cline{2-5}
& \textbf{DD} & \textbf{NCI-1} & \textbf{NCI-109} &\textbf{Mutagenicity}\\
\toprule
0.25 & $76.84 \pm 1.55$ & $67.06\pm2.38$ & $74.35 \pm 0.99$ & $76.49 \pm 0.49$ \\
0.5 & \bm{$77.31 \pm 1.54$} & \bm{$69.14 \pm 1.45$} & \bm{$76.31 \pm 1.22$} & \bm{$78.28 \pm 1.35$}\\
0.75 & $77.10 \pm 0.98$ & $68.48 \pm 1.86$ & $75.90\pm1.13$&$77.41\pm1.33$\\
\bottomrule
\par
\end{tabular}
\label{table2-1}
\end{table*}
\begin{table*}
\caption{The evaluation of different weight $\alpha$ based on the ${GSAPool}_{GCNConv\&MLP+{Fusion}_{GAT}}$ architecture.}
\begin{tabular}{c|cccc}
\toprule
\multirow{2}{*}{\textbf{Weight} \bm{$\alpha$}} & \multicolumn{4}{c}{\textbf{Datasets}}\\
\cline{2-5}
& \textbf{DD} & \textbf{NCI-1} & \textbf{NCI-109} &\textbf{Mutagenicity}\\
\toprule
0 & $75.92\pm1.13$ &$72.84\pm1.89$ & $73.02\pm2.15$ &$78.17\pm1.16$ \\
0.2 & $77.26 \pm 1.85$ & $74.87 \pm 1.32$ & $75.73\pm 2.06$ & $80.13\pm1.15$\\
0.4 & $78.40 \pm 1.33$ & \bm{$75.20 \pm 1.55$}& \bm{$77.43 \pm 1.35$} & \bm{$81.99 \pm 1.20$}\\
0.6 & \bm{$80.84 \pm 1.17$} & $72.77 \pm 1.50$ & $77.15\pm1.10$ & $79.24\pm0.89$\\
0.8 & $79.28 \pm 0.85$ & $71.78 \pm1.39$ &$76.83\pm 1.88$ & $78.73\pm1.14$ \\
1 & $77.31 \pm 1.54$ & $69.14 \pm 1.45$& $76.31 \pm 1.22$ & $78.28 \pm 1.35$\\
\bottomrule
\end{tabular}
\label{table2-2}
\end{table*}

\begin{table*}[h]
\caption{Performance of different node evaluation functions used in the proposed GSAPool.}
\begin{tabular}{l|cc|cccc|c}
\toprule
 \multirow{2}{*}{\textbf{Evaluation Functions}}& \multicolumn{2}{c|}{\textbf{Type}} & \multicolumn{4}{c|}{\textbf{Datasets}} & \multirow{2}{*}{\textbf{Average Accuracy}}\\
\cline{2-7}
& Structure & Feature & DD & NCI-1 & NCI-109 & Mutagenicity & \\
\toprule
GraphSAGE\cite{WLH17} & $\times$ & \checkmark & $72.60 \pm 3.84$ & $65.93 \pm 2.10$ & $70.32 \pm 1.77$ & $70.19 \pm 3.10$ & 69.76\\
GAT\cite{PV18} & $\times$ & \checkmark & $75.37 \pm 1.09$ & $64.61 \pm 0.88$ & $67.85 \pm 2.10$ & $75.56 \pm 1.20$ & 70.85\\
GCNConv\cite{TN17} & \checkmark & \checkmark & $77.05 \pm 0.75$ & $61.05 \pm 0.28$ &  $70.18 \pm 4.32$ & $76.14 \pm 1.58$ & 71.11\\
ChebConv\cite{MD16} & \checkmark & \checkmark & $77.35 \pm 1.39$ & $71.45 \pm 1.92$ & $74.26 \pm 2.14$ &$77.65 \pm 1.24$ & 75.18\\
MLP\cite{JJH82} & $\times$ & \checkmark & $74.03 \pm 1.32$ & $72.60 \pm 1.38$ & $75.55 \pm 1.58$ & $79.12 \pm 1.05$ & 75.33\\
GCNConv\&MLP & \checkmark & \checkmark & \bm{$78.27 \pm 0.96$} & \bm{$72.84 \pm 1.15$} & \bm{$77.07 \pm 1.89$} & \bm{$79.90 \pm 0.80$} & \textbf{77.02}\\
\bottomrule
\end{tabular}
\label{table3}
\end{table*}
\begin{table*}[h]
\caption{Performance of different feature fusion functions used in GSAPool.}
\begin{tabular}{l|cccc|c}
\toprule
\textbf{Fusion Functions} & \textbf{DD} & \textbf{NCI-1} & \textbf{NCI-109} & \textbf{Mutagenicity} & \textbf{Average}\\
\toprule
No Fusion & $77.05 \pm 0.75$ & $61.05 \pm 0.28$ &  $70.18 \pm 4.32$ & $76.14 \pm 1.58$ & 71.11\\
GCNConv-based Fusion &  \bm{$77.94 \pm 2.00$} & $67.95 \pm 1.91$ & $72.98 \pm 1.85$ & $76.89 \pm 1.12$ & 73.94\\
GAT-based Fusion & $77.31 \pm 1.54$ & \bm{$69.14 \pm 1.45$} & \bm{$76.31 \pm 1.22$} & \bm{$78.28 \pm 1.35$} & \textbf{75.26}\\
\bottomrule
\end{tabular}
\label{table4}
\end{table*}
\begin{table*}[h]
\caption{Comparison with the state-of-the-art graph pooling methods.}
\begin{tabular}{l|cccc|c}
\toprule
\textbf{Methods} & \textbf{DD} & \textbf{NCI-1} & \textbf{NCI-109} & \textbf{Mutagenicity} & \textbf{Average}\\
\toprule
gPool\cite{GH19} & $75.01 \pm 0.86$ & $67.02 \pm 2.25$ & $66.12 \pm 1.60$ & $67.44 \pm 2.78$ & 68.90\\
SAGPool\cite{JL19} & $76.45 \pm 0.97$ & $67.45 \pm 1.11$ & $67.86 \pm 1.41$ & $76.89 \pm 1.12$ & 72.16\\
SET2SET\cite{VO15} & 74.50 & 71.50 & 68.60 & 76.40 & 72.75\\
DiffPool\cite{RY18} & 78.00 & 76.00 &  74.10 & 80.60 & 77.18\\
EigenPooling\cite{YM19} & 78.60 & \textbf{77.00} & 74.90 & 79.50 & 77.50\\
\midrule
${GSAPool}_{GCNConv\&MLP}$ & $78.27 \pm 0.96$ & $72.84 \pm 1.15$ & $77.07 \pm 1.89$& $79.90 \pm 0.80$ & 77.02\\
${GSAPool}_{GCNConv+{Fusion}_{GAT}}$ & $77.31 \pm 2.00$ & $69.14 \pm 1.45$ & $76.31 \pm 1.22$ & $78.28 \pm 1.35$ & 75.26\\
${GSAPool}_{GCNConv\&MLP+{Fusion}_{GAT}}$ & \bm{$80.84 \pm 1.17$} & $75.20 \pm 1.55$ & \bm{$77.43 \pm 1.35$} & \bm{$81.99 \pm 1.20$} & \textbf{78.87}\\
\bottomrule
\end{tabular}
\label{table5}
\end{table*}
\begin{figure*}[h]
\center{\includegraphics[width=14cm,height=4cm]{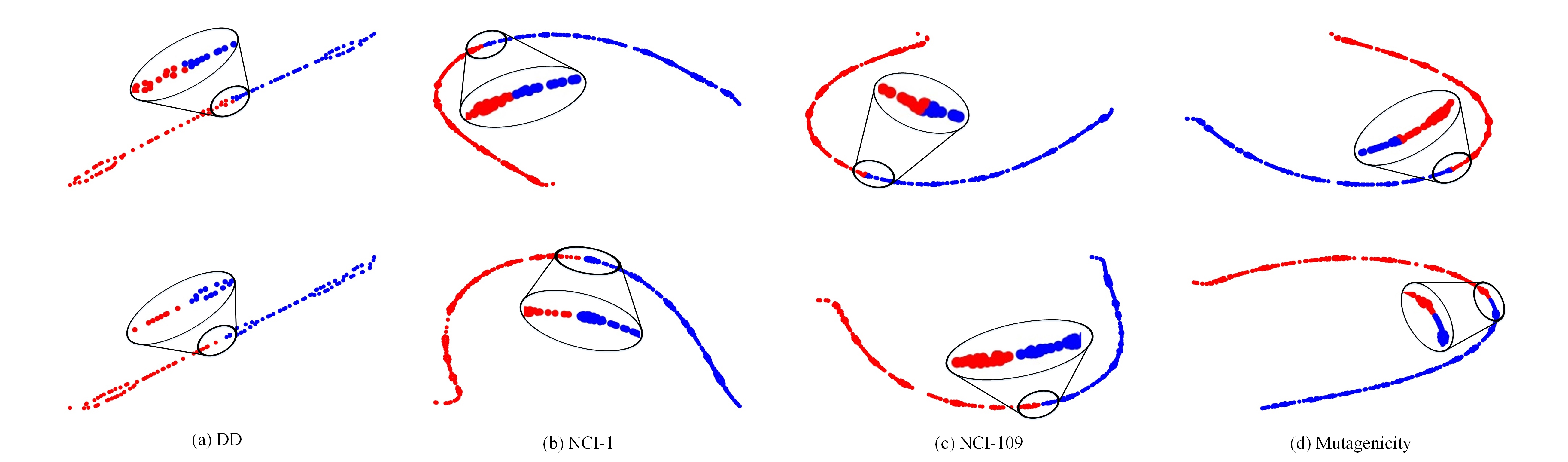}}
\caption{The t-SNE \cite{LM08} visualization of the features of the pooling graph results on DD, NCI-1, NCI-109, and Mutagenicity. The features of the proposed GSAPool (in the second line) are more distinguishable than the features of SAGPool \cite{JL19} (in the first line). }
\label{experFig}
\end{figure*}
\subsection{Training Configurations}
In our experiments, we uses 10-fold cross validation to evaluate the pooling methods. And Nvidia Titan V is the version of GPU used in our experiments. All the GSAPool models are implemented with PyTorch\cite{PA17} and the geometric deep learning extension library\cite{FM19}.
\subsection{Ablation Study}
\textbf{Hyperparameter Analysis}: Two hyperparameters are used in our experiments: pooling ratio and combination weight $\alpha$. For the pooling ratio, there are three value used: 0.25, 0.5 and 0.75. When the pooling ratio is 0.25, the pooling method performs not so well because the selected feature information is little. And the performance is not good when pooling ratio is 0.75, for the graph retaining too much redundant information. Therefore, 0.5 is finally taken as the pooling ratio. $\alpha$ is an empirical value, and several values are tested to find the best. Finally, we find that when the dataset is DD \cite{DPD13}\cite{SN11}, the best value of $\alpha$ is 0.6, and for the other three datasets, 0.4 is the best value. Table \ref{table2-1} and Table \ref{table2-2} show all the hyperparameters used in this study, respectively.\\
\textbf{Topology Learning Strategy Analysis}: We compare GCNConv\\ \cite{TN17}, ChebConv \cite{MD16}, GraphSAGE \cite{WLH17} and GAT \cite{PV18} in GSAPool as graph pooling topology learning methods on four datasets (DD \cite{DPD13}\cite{SN11}, NCI-1 \cite{WN08}\cite{SN11}, NCI-109 \cite{WN08}\cite{SN11}, and Mutagenicity \cite{JK05}\cite{KR08}). The average accuracies of graph classification and corresponding standard deviations are expressed in percentages. Table \ref{table3} shows that GCNConv and ChebConv are better than GraphSAGE and GAT. MLP \cite{JJH82} is excellent on NCI-1, NCI-109, and Mutagenicity. 

GCNConv and ChebConv consider structural information when calculating the node importance score, whereas GAT and SAGE do not. Therefore, GCNConv can use additional graph information to create more representative graph vectors. As a result, GCNs perform best on graph classification task. MLP performs well on NCI-1 , NCI-109 and Mutagenicity but not so well on DD. For example, the average number of nodes and edges in each graph of DD is more than that in NCI-1, which means that the graph structure in DD is highly complex. MLP focuses on feature information rather than structural information, so it can get better performance in NCI-1, of which the graph structure is simpler than that of DD. NCI-109 and Mutagenicity are similar. However, when combining GCNConv and other GNNs, such as GAT, the result is not as good as the combination with MLP. From our perspective, GCNConv is more similar with GAT than MLP; thus, the area of overlap in the learning space of GCNConv and GAT is larger than that of GCNConv and MLP. Therefore, the combination of GCNConv and MLP can learn more features than that of GCNConv and GAT, which is why MLP is better than GAT as a feature mode selection.\\
\textbf{Feature Fusion Strategy Analysis}: The GCNConv-based \cite{TN17} fusion performs well on DD \cite{DPD13}\cite{SN11}, while GAT-based fusion \cite{PV18} performs well on NCI-1 \cite{WN08}\cite{SN11}, NCI-109 \cite{WN08}\cite{SN11} and Mutagenicity \cite{JK05}\cite{KR08} (see Table \ref{table4}). The reason for the success of the fusion mechanism is that the use rate of the node feature information increases. As a result, the fusion mechanism creates a highly representative graph vector; thus, it has a high accuracy on graph classification. In addition, GCNConv considers the structural information when aggregating the information of the nodes while structure is not considered by GAT. Thus, the GCNConv performs well highly intricate structure while GAT is effective on the simple datasets.
\subsection{State-of-the-art}
Table \ref{table5} shows that graph pooling topology generation and node feature fusion measure perform better when combined than when each method is used alone. Moreover, our proposed method is higher than the best one of the current pooling methods on DD \cite{DPD13}\cite{SN11}, NCI-109 \cite{WN08}\cite{SN11} and Mutagenicity \cite{JK05}\cite{KR08}. Our method is not as good as DiffPool \cite{RY18} and EigenPooling \cite{YM19} only on NCI-1 \cite{WN08}\cite{SN11}. However, our method has the highest average accuracy. Table \ref{table5} demonstrates that our method reaches state-of-the-art level. Figure \ref{experFig} shows the t-SNE \cite{LM08} visual comparison results of SAGPool \cite{JL19} and the proposed GSAPool on the graph classification task. Graphs can be easily separated by a horizontal or vertical line and the gap between the two categories is highly evident in GSAPool.

Compared with DiffPool and EigenPooling, the proposed GSAPool used structure and feature information explicitly at the same time. Therefore, GSAPool can use additional graph information to construct the pooled graph. GSAPool outperforms gPool and SAGPool because GSAPool considers the information of the node synthetically and uses feature fusion mechanism, which enables the pooled graph to represent the original graph reasonably and accurately. Thus, GSAPool can obtain an improved performance.
\section{Conclusion}
In this study, we proposed GSAPool to solve the problems in the top-k selection graph pooling. On the one hand, our method used three graph pooling topology generation strategies. On the other hand, node feature fusion was adopted before the unimportant nodes were discarded so that the node information could be used efficiently. Lastly, we combined the two approaches and compared them with the current approaches. The experimental results demonstrated that our proposed method achieved state-of-the-art graph classification performance on the benchmark datasets. Our GSAPool method can be easily integrated into other deep GNN architectures. 
\section{Acknowledgments}
This research is supported by National key research and development plan (2019YFB1311600) and Ningbo 2025 Key Project of Science and Technology Innovation (2018B10071)


\bibliographystyle{ACM-Reference-Format}
\bibliography{Reference}

\end{document}